%                                                                  %
% This file contains corrections made on thursday January 16, 1997 %
%                                                                  %
%                                                                  %
%%%%%%%This requires the PHYZZX.TEX macropackage

%%%%%%%If you do not have the msbm fonts, delete the following 4 lines

%%%%%%%%%%%%

%%%%%%%%%%

\tolerance=10000
\input phyzzx
\input epsf.tex
 \def\unit{\hbox to 3.3pt{\hskip1.3pt \vrule height 7pt width 
.4pt \hskip.7pt
\vrule height 7.85pt width .4pt \kern-2.4pt
\hrulefill \kern-3pt
\raise 4pt\hbox{\char'40}}}

\def\tr{\rm tr}

%%%%%%%%%%%%%%%%%%%%%%%%%%%%%%%%%%%%%%%%%%%%%%%%%%%%%%%%%%%%%%%%%%%%%%%%%%%%%
\REF\pol{J. Polchinski, Phys. Rev. Lett. {\bf 75} (1995) 4724.}
\REF\leigh{R.G. Leigh, Mod. Phys. Lett. {\bf A 4} (1989) 2767.}
\REF\Li{M. Li, Nucl. Phys. {\bf B460} (1996) 351; C.G. Callan and I.R. Klebanov, Nucl. Phys. {\bf 443} (1995) 444.}
\REF\doug{M. Douglas, {\it Branes within branes}, hep-th/9512077.}
\REF\bst{E. Bergshoeff, E. Sezgin and P.K. Townsend, Phys. Lett. {\bf 189B}
(1987) 75; Ann. Phys. (NY) {\bf 185} (1988) 330.}
\REF\pkt{P.K. Townsend, Phys. Lett. {\bf B373} (1996) 68.}
\REF\ceder{M. Cederwall, A. von Gussich, B.E.W. Nilsson and A. Westerberg, {\it
The Dirichlet super-three-brane in type IIB supergravity}, hep-th/9610148.}
\REF\schwarz{M. Aganagic, C. Popescu and J.H. Schwarz, {\it D-brane actions
with local kappa-symmetry}, hep-th/9610249.}
\REF\howe{P.S. Howe and E. Sezgin, {\it Superbranes}, hep-th/9607227.}
\REF\sch{C. Schmidhuber, Nucl. Phys. {\bf B467} (1996) 146.}
\REF\tsey{A. Tseytlin, Nucl. Phys. {\bf B469} (1996) 51.}
\REF\sbi{E. Bergshoeff, M. Rakowski and E. Sezgin, Phys. Lett. {\bf B185}
(1987) 371; S. Cecotti and S. Ferrara, Phys. Lett. {\bf B187} (1987) 335; A.
Karlhede, U. Lindstrom, M. Rocek and G. Theodorides, Nucl. Phys. {\bf B294}
(1987) 498; J. Bagger and A. Galperin, {\it A new Goldstone multiplet for
partially broken supersymmetry}, hep-th/9608177}
\REF\gates{S.J. Gates Jr., Nucl. Phys. {\bf B184} (1981) 381.}
\REF\bdr{E. Bergshoeff and M. de Roo, Phys. Lett. {\bf B380} (1996) 265.}
\REF\ght{M.B. Green, C.M. Hull and P.K. Townsend, Phys. Lett. {\bf 382B} (1996)
65.}
\REF\pktb{P.K. Townsend, {\sl Three lectures on supermembranes}, in  {\it
Superstrings '88}, eds. M. Green, M. Grisaru, R.  Iengo, E. Sezgin and A.
Strominger, (World Scientific 1989).}
\REF\DHIS{J.L. Carr, S.J. Gates, Jr., and R.N. Oerter, Phys. Lett. {\bf
189B} (1987) 68; S. Bellucci, S. Gates, B. Radak, P. Majumdar and
S. Vashakidze, Mod. Phys. Lett. {\bf A21} (1989) 1985.}
\REF\nilsson{B.E.W. Nilsson, Nucl. Phys. {\bf B188} (1981) 176.}
\REF\howewest{P.S. Howe and P.C. West, Nucl. Phys. {\bf B238} (1984) 181.}
\REF\ant{A. Aurilia, H. Nicolai and P.K. Townsend, Nucl. Phys. {\bf B176} 
(1980) 509.}
\REF\cedertwo{M. Cederwall, A. von Gussich, B.E.W. Nilsson, P. Sundell, and A.
Westerberg, {\it The Dirichlet super-p-branes in ten-dimensional Type IIA and
IIB supergravity}, hep-th/9611159.}

%%%%%%%%%%%%%%%%%%%%%%%%%%%%%%%%%%%%%%%%%%%%%%%%%%%%%%%%%%%%%%%%%%%%

\Pubnum{ \vbox{ \hbox{R/96/53} \hbox{UG-8/96} \hbox{hep-th/9611173}}}
\pubtype{}
\date{November 1996}

\titlepage

\title {\bf Super D-branes}

\author{E. Bergshoeff}
\address{Institute for Theoretical Physics, Nijenborgh 4,
\break
9747 AG Groningen, The Netherlands}
\andauthor{P.K. Townsend}
\address{DAMTP, University of Cambridge,
\break
Silver St., Cambridge CB3 9EW, U.K.}

\abstract{We present a manifestly Lorentz invariant, spacetime supersymmetric,
and `$\kappa$-invariant' worldvolume action for all type II Dirichlet p-branes,
$p\le9$, in a general type II supergravity background, including massive
backgrounds in the IIA case. The $p=0,2$ cases are rederived from D=11. 
The $p=9$ case provides a supersymmetrization of the D=10 Born-Infeld action.}

\endpage
%\pagenumber=1

%%%%%%%%%%%%%%%%%%%%%%%%%%%%%%%%%CHAPTER 1%%%%%%%%%%%%%%%%%%%%%%%%

\chapter{Introduction}

The worldvolume actions for p-brane solutions of supersymmetric field theories
may be viewed as $(p+1)$-dimensional non-linear sigma-models with a superspace
as the target space, i.e. the worldvolume fields $Z^M(\xi)$ define a map
$W\rightarrow \Sigma$ from the worldvolume $W$ with coordinates $\xi^i$
($i=0,1,\dots, p$) to a superspace $\Sigma$ with coordinates
$Z^M=(x^m,\theta^\mu)$. Many p-brane solutions of
supergravity theories do not quite fit this pattern, however, because their
worldvolume fields include vectors or antisymmetric tensors. Examples are
provided by the Ramond-Ramond (R-R) p-branes of ten-dimensional (D=10) type II
supergravity theories.  Because these have an interpretation as Dirichlet
p-branes, or D-p-branes, of type II superstring theory [\pol], their bosonic
worldvolume actions (to leading order in $\alpha'$) can be computed by standard
superstring methods [\leigh,\Li,\doug]. 
The worldvolume action includes a 1-form
gauge potential $V$, the so-called Born-Infeld (BI) field, which couples to the
endpoints of type II strings `on the brane'. For reasons explained elsewhere
(see, e.g.  [\bst]) the D=10 Lorentz covariant form of this action must
have a fermionic gauge invariance, usually called `kappa-symmetry'. Given the
bosonic D-brane action, it is not difficult to guess the form of the super
D-brane action, even in a general supergravity background, but the complications
due to the BI gauge field have so far prevented the construction of the complete
$\kappa$-symmetry transformations for general $p$, and the verification of
$\kappa$-invariance, although a number of partial results have been obtained. A
form of the super D-2-brane action and its $\kappa$-symmetry transformations, in
a Minkowski background, was deduced from that of the D=11 supermembrane by
means of IIA/M-theory duality [\pkt]. More recently, the super D-3-brane action
and its $\kappa$-symmetry in a general IIB supergravity background has been
presented [\ceder]. Also, an action for general $p$ in a Minkowski background
has been given and a strategy for verifying its $\kappa$-symmetry proposed
[\schwarz]. We should also mention that super D-p-brane actions with simultaneous
worldvolume and spacetime supersymmetry have been found [\howe], but the
relation to the Green-Schwarz-type action considered here has not yet been
spelled out. 

Here, we  present a Lorentz invariant and supersymmetric worldvolume
action for all type II Dirichlet p-branes, $p\le9$, in a general type II
supergravity background. We also give the explicit form of the $\kappa$-symmetry
transformations, which we have fully verified in bosonic backgrounds for
$p\le6$ and partially verified for $p>6$. We also show how the $p=0,2$ actions
can be obtained from D=11. The $p=0$ action is obtained by reduction of the D=11
massless superparticle. The
$p=2$ action is obtained by reduction of the D=11 supermembrane action, followed
by a scalar/vector duality transformation, as in [\pkt] but incorporating the
results of [\sch,\tsey] so as to arrive at the standard Born-Infeld form of the
action. The $\kappa$-symmetry transformations of the super D-2-brane
action can also be deduced from D=11, but the form of these transformations do
not obviously generalize to higher $p$. A redefinition of the $\kappa$-symmetry
parameter is needed to put them in the form used in this paper.  

Although the focus of this paper is on the D-p-branes for $p\le8$, our
main result applies equally when $p=9$. This special case is equivalent to a
supersymmetrization of the D=10 Born-Infeld action, which was only partially
known previously [\sbi]. Thus, subject to a full verification of
$\kappa$-symmetry for $p=9$, this problem is now solved, at least in principle. 
A curious feature of this approach to the super Born-Infeld action is that it
makes essential use of an 11-form superspace field strength, which vanishes
identically when restricted to spacetime! The
possibility of such superspace gauge fields has been explored in the past
[\gates] and our work provides a nice example of their utility.

The organization of the remainder of this paper is as follows. We first give
the super D-p-brane action and explain some of the superspace conventions.
We then present some preliminary results needed to compute the
$\kappa$-variation of this action. We then perform the calculation and verify
$\kappa$-symmetry in detail for $p\le6$ and partially for $p=7,8,9$. We then
rederive the $p=0,2$ results from known results in D=11, including the
generalization to fermionic and `massive' backgrounds. 

%%%%%%%%%%%%%%%%%%%%%%%%%%%%%%%%%CHAPTER 2%%%%%%%%%%%%%%%%%%%%%%%%

\chapter{The super D-brane action}

Our proposed super D-brane action, in a general N=2 supergravity background,
(for the string-frame metric) takes the form
$$
I= I_{DBI} + I_{WZ} 
\eqn\introa
$$
where 
$$
I_{DBI}= -\int d^{p+1}\xi\, e^{-\phi}\sqrt{-\det (g_{ij} + {\cal F}_{ij})}
\eqn\introb
$$
is a Dirac-Born-Infeld type action and $I_{WZ}$ is a Wess-Zumino (WZ) type 
action to be discussed below; ${\cal F}_{ij}$ are the components of a
`modified' 2-form field strength 
$$
{\cal F} = F-B\ ,
\eqn\introc
$$
where $F=dV$ is the usual field strength 2-form of the BI field $V$ and $B$ is
the pullback to the worldvolume of a 2-form potential $B$ on superspace, whose
leading component in a $\theta$-expansion is the 2-form potential of
Neveu-Schwarz/Neveu-Schwarz (NS-NS) origin in type II superstring theory. We
use the same letter for superspace forms and their pullbacks to the
worldvolume since it should be clear from the context which is meant.
Superspace forms may be expanded on the coordinate basis of 1-forms $dZ^M$ or
on the inertial frame basis $E^A= dZ^ME_M{}^A$, where $E_M{}^A$ is the
supervielbein. The basis $E^A$ decomposes under the action of the Lorentz group
into a Lorentz vector $E^a$ and a Lorentz spinor. The latter is a 32-component
Majorana spinor for IIA superspace and a pair of chiral Majorana spinors for
IIB superspace. Thus
$$
E^A =\cases{(E^a,E^\alpha) & (IIA)\cr
(E^a, E^{\alpha\; I})\, I=1,2 & (IIB)}
\eqn\introca
$$
In the IIB case we still allow $\alpha$ to run from 1 to 32 but include a
chiral projector as appropriate. The worldvolume
metric $g_{ij}$ appearing in \introb\ is defined in the standard way as 
$$
g_{ij} = E_i^a E_j^b\eta_{ab}
\eqn\introd
$$
where $\eta$ is the D=10 Minkowski metric and
$$
E_i^A = \partial_i Z^M E_M{}^A\ .
\eqn\introe
$$

Thus $I_{DBI}$ is a straightforward extension to superspace of the corresponding
term in the bosonic action. The same is true for the WZ term except for
one new feature of relevance to the 9-brane. We introduce a R-R potential $C$
as a formal sum of $r$-form superspace potentials $C^{(r)}$, i.e.  
$$
C = \sum_{r=0}^{10} C^{(r)}\ .
\eqn\introf
$$
The even potentials are those of IIB supergravity while the odd ones are those
of IIA supergravity. In the bosonic case one could omit the 10-form gauge
potential $C^{(10)}$ on the grounds that its 11-form field strength is
identically zero. But an 11-form field strength on {\it superspace} is not
identically zero; in fact we shall see that it is non-zero even
in a Minkowski background, a fact that is crucial to the $\kappa$-symmetry of
the super 9-brane action.

The WZ term can now be written as [\Li,\doug,\bdr,\ght]
$$
I_{WZ} =\int_W C e^{\cal F}\  + \ mI_{CS}
\eqn\introg
$$
where, in the first term, the product is understood to be the exterior product of
forms and the form of appropriate degree is chosen in the the `form-expansion'
of the integrand, i.e. (p+1) for a D-p-brane. The $I_{CS}$ term is a
(p+1)-form Chern-Simons (CS) action that is present (for odd $p$) in a {\it
massive} IIA background; its coefficient $m$ is the IIA mass parameter.
This WZ term is formally the same as the known {\it bosonic} D-brane WZ action,
but  here the forms $C^{(r)}$ and $B$ are taken to be forms on {\it superspace},
e.g.
$$
C^{(r)} = {1\over r!} dZ^{M_1}\cdots dZ^{M_r} C_{M_r\dots M_1}\ .
\eqn\introh
$$
This illustrates the standard normalization and the `reverse order' convention
for components of superspace forms. This convention goes hand in hand with the
convention for exterior differentiation of superspace forms in which the exterior
derivative acts `from the right'. Thus,
$$
dC^{(r)} = {1\over r!}dZ^{M_1}\cdots dZ^{M_r} dZ^N\partial_N C_{M_r\dots M_1}
\eqn\introi
$$

As explained in [\ght], the field strength for the RR field $C$ is
$$
R(C) = dC - HC + me^B
\eqn\introj
$$
where $m$ is the mass parameter of the IIA theory. $R(C)$ can be
written as the formal sum
$$
R(C) = \sum_{n=1}^{11} R^{(n)}\ .
\eqn\introk
$$
Note that the top form is an 11-form because we included a 10-form $C^{(10)}$ in
the definition of $C$. The field strengths $R^{(n)}$ will be subject to
superspace constraints, to be given below for bosonic backgrounds, in addition
to the constraint described in [\ght] relating the bosonic components of
$R^{(n)}$ to the Hodge dual of the bosonic components of $R^{(10-n)}$. 
This concludes our presentation of the super D-brane action. 

%%%%%%%%%%%%%%%%%%%%%%%%%%%%%Chapter 3%%%%%%%%%%%%%%%%%%%%%%%%%%%%

\chapter{$\kappa$-symmetry preliminaries}

Given variations $\delta Z^M$ of the worldvolume fields $Z^M$, we define
$$
\eqalign{
\delta E^A &\equiv \delta Z^M E_M{}^A \cr
\delta \Omega_B{}^A &= \delta Z^M \Omega_{M\, B}{}^A }
\eqn\prelima
$$
where $\Omega$ is the connection on the tangent bundle of superspace. As usual,
we take the structure group of the tangent bundle to be the Lorentz group so
that, in particular,
$$\eqalign{
\Omega_b{}^\alpha &=\Omega_\beta{}^a =0\cr
\Omega_\alpha{}^\alpha &= \Omega_{(b}{}^a\eta_{c)a}=0 \ .}
\eqn\prelimb
$$
A useful lemma is 
$$
\delta E_i^A = \partial_i(\delta E^A) -\delta E^B E_i^C T_{CB}{}^A 
-E_i^B\delta\Omega_B{}^A + \delta E^B \partial_i Z^N \Omega_{N\, B}{}^A 
\eqn\prelimc
$$
where $T^A$ is the torsion 2-form, i.e. $T^A= {1\over2}E^B\wedge E^C
T_{BC}{}^A$. This corrects a result quoted in [\bst,\pktb] in which the last term
was missing.

The IIA torsion tensor is subject to the constraints [\DHIS]
$$
\eqalign{
T_{\beta\gamma}{}^a &= i\Gamma^a_{\alpha\beta}\cr
T_{b\, \gamma}{}^a & = \delta_a^b \chi_\gamma }
\eqn\prelimd
$$
where $\chi$ is a spinor proportional to the dilatino of the
supergravity background, i.e. $\chi\propto D\phi\,$ [\nilsson]. The IIB torsion
tension is subject to the constraints [\howewest]
$$
\eqalign{
T_{\alpha\, I\; \beta\, J}{}^a &=  i\delta_{IJ} \big(\Gamma^a 
{\cal P}\big)_{\alpha\beta}\cr
T_{b\, \gamma\, K}{}^a & = \delta_a^b \chi_{\gamma\, K} }
\eqn\prelime
$$
where ${\cal P}$ is a chiral projection operator and $\chi^I$ is a pair of
chiral spinors, i.e. ${\cal P}\chi^I =\chi^I$.

A universal feature of $\kappa$-symmetry is that
$$
\delta_\kappa E^a =0\ .
\eqn\prelimf 
$$ 
Using this fact in the lemma \prelimc, and taking into account the
restrictions \prelimb\ on the form of $\Omega$, we have that
$$
\delta_\kappa g_{ij} = -2 E_i^a\delta_\kappa E^B E_j^C T_{CB}{}^b \eta_{ab}\ .
\eqn\prelimg
$$
Using \prelimf\ again, and the torsion constraints, we find that
$$
\delta_\kappa g_{ij} = \cases{ -2i \delta_\kappa E\Gamma_{11}\gamma_{(i} E_{j)} 
- 2 g_{ij} \delta_\kappa E\chi & (IIA)\cr
 -2i \delta_\kappa E^I\gamma_{(i} E_{j)}^I 
- 2 g_{ij} \delta_\kappa E^I\chi^I & (IIB)\ .}
\eqn\prelimi
$$
Here, $\delta_\kappa E$ and $\delta_\kappa E^I$ are the spinor components of
$\delta_\kappa E^A$ for IIA and IIB, respectively, the vector components
vanishing by hypothesis, and $E_i$ and $E_i^I$ are the spinor components of
$E_i^A$. Since all IIB spinors are chiral, the chirality projection operator ${\cal P}$ may be omitted in the IIB expression. We have also supressed spinor indices. The matrices $\gamma_i$ are defined
as
$$
\gamma_i = E_i{}^a\Gamma_a\ .
\eqn\prelimj
$$
The round brackets enclosing indices indicate symmetrization with `strength
one'; we use square brackets for antisymmetrization. We also adopt the
convention by which spinor indices are raised and lowered with the charge
conjugation matrix $C$, as explained e.g. in [\pktb]. Thus
$\Gamma^a_{\alpha\beta}$ are the components of the symmetric matrix formed from
the product $C\Gamma^a$.

Another useful lemma concerns the variation of a (p+1)-form A induced by the
variation $\delta Z^M$. Defining $F(A)=dA$, we have
$$
\eqalign{
\delta A = {1\over (p+1)!} &E^{A_{p+2}} \cdots E^{A_2} \delta E^{A_1}
F(A)_{A_1A_2\dots A_{p+2}}\cr
&\qquad  + {1\over p!} d\bigg( E^{A_p} \dots E^{A_1}\delta
E^B A_{BA_1\dots A_p}\bigg) \ .}
\eqn\prelimk
$$
This corrects a result quoted in [\pktb] in which the total derivative term is
missing. An application of this result to the NS-NS 2-form $B$ with field
strength $H=dB$ leads to
$$
\delta B = {1\over2}E^A E^B\delta E^C H_{CBA} + 
d\big(E^A \delta E^B B_{BA}\big)\ .
\eqn\preliml
$$
One useful piece of information that one learns from the D=11 origin of the
D-2-brane is that
$$
\delta_\kappa V_i = E_i^A\delta_\kappa E^B B_{BA}
\eqn\prelimm
$$
so that 
$$
\delta_\kappa {\cal F}_{ij} =  -E_{[i}^A E_{j]}^B\delta_\kappa E^C H_{CBA}\ .
\eqn\prelimn
$$
A feature of this result is that ${\cal F}$ is `$\kappa$-covariant' in the
sense that its $\kappa$-variation does not involve derivatives of $\kappa$,
a property that is crucial for $\kappa$-invariance. 

The IIA superspace constraints on $H$ are [\DHIS]
$$
\eqalign{
H_{\alpha\beta\gamma} &=0\cr
H_{\alpha\beta\, c} &=-i\big(\Gamma_{11}\Gamma_a\big)_{\alpha\beta}
\cr H_{\alpha\, bc} &= (\Gamma_{bc}\zeta)_\alpha }
\eqn\prelimo
$$
where $\zeta$ is another spinor proportional to $D\phi$. The IIB
superspace constraints on $H$ are [\DHIS]
$$
\eqalign{
H_{\alpha\, I\, \beta\, J\, \gamma\, K} &=0\cr
H_{\alpha\, I\, \beta\, J\, c} &=i(\sigma_3)_{IJ}\big(\Gamma_a 
{\cal P}\big)_{\alpha\beta}\cr
H_{\alpha\, I\, bc} &= (\Gamma_{bc}\zeta)_\alpha^I\ . }
\eqn\prelimp
$$

If these constraints are now used in \prelimn\ we
have, for IIA,
$$
\delta_\kappa {\cal F}_{ij} =  -2i\delta_\kappa E \Gamma_{11}\gamma_{[i} E_{j]}
+ \delta_\kappa E \gamma_{ij}\zeta
\eqn\prelimr
$$
and for IIB,
$$
\delta_\kappa {\cal F}_{ij} =  2is_{IJ}\delta_\kappa E^I \gamma_{[i} E_{j]}^J +
\delta_\kappa E^I \gamma_{ij}\zeta^I
\eqn\prelims
$$

It remains to give the superspace constraints on the R-R field strengths.
These constraints imply that all components of $R^{(n)}$ with more than two
spinor indices vanish. The components with no spinor indices are unconstrained
apart from the Hodge dual relation between the components of $R^{(n)}$ and 
$R^{(10-n)}$. The components
with one spinor index are all proportional to the dilatino and hence vanish
for bosonic backgrounds. The only other non-vanishing components are those with
precisely two spinor indices; the constraints on these components are needed
for verification of $\kappa$-symmetry. They are 
$$
R^{(n+2)}_{\alpha\beta a_1\dots a_n} =\cases{ie^{-\phi}
(\Gamma_{a_1\dots a_n}\Gamma_{11})_{\alpha\beta} & (n=0,4,8) \cr
ie^{-\phi}(\Gamma_{a_1\dots a_n})_{\alpha\beta} & (n=2,6)}
\eqn\twoacon
$$
for the IIA R-R field strengths, while
$$
R^{(n+2)}_{\alpha\, I\,\beta\, J\, a_1\dots a_n}
=\cases{ie^{-\phi}(\Gamma_{a_1\dots a_n}{\cal P})_{\alpha\beta}
(\sigma_1)_{IJ} & (n=1,5,9)\cr 
ie^{-\phi}(\Gamma_{a_1\dots
a_n}{\cal P})_{\alpha\beta}(i\sigma_2)_{IJ} & (n=3,7) }
\eqn\twobcon
$$
for the IIB R-R field strengths.

Finally, It is convenient to introduce the matrix
$$
\Gamma_{(0)} = {1\over (p+1)!\sqrt{-\det g}}\, \varepsilon^{i_1\dots
i_{(p+1)}}\gamma_{i_1\dots i_{(p+1)}}
\eqn\gamzero
$$
which has the properties
$$
(\Gamma_{(0)})^2 = (-1)^{p(p-1)} \qquad
\Gamma_{(0)}\gamma^i 
=(-1)^p\gamma^i\Gamma_{(0)}\ .
\eqn\sqgam
$$
A useful identity involving $\Gamma_{(0)}$ is
$$
\varepsilon^{i_1\dots i_kj_{k+1}\dots j_{p+1}} \gamma_{j_{k+1}\dots j_{p+1}} =
(-1)^{k(k-1)/2}(p-k)!\sqrt{-\det g}\, \gamma^{i_i\dots i_k}\, \Gamma_{(0)}\, .
\eqn\idengam
$$

%%%%%%%%%%%%%%%%%%%%%%%%%%%%%Chapter 4%%%%%%%%%%%%%%%%%%%%%%%%%%%%

\chapter{Proof of $\kappa$-invariance}

We are now in a position to compute the $\kappa$-variation of the proposed
super D-p-brane action in a bosonic background. We begin with the variation of
the DBI term, for which the Lagrangian is
$$
{\cal L}_{DBI} = -e^{-\phi}\sqrt{-\det(g+{\cal F})}\ .
\eqn\dbidef
$$
Our results for $\delta_\kappa {\cal L}_{DBI}$ are essentially the
same as those obtained in [\ceder,\schwarz], but we need them in our 
conventions. Using the results of the previous section for the
$\kappa$-variations of $g_{ij}$ and ${\cal F}_{ij}$, but now omitting terms which
vanish in a bosonic background, we have
$$
\delta_\kappa (g_{ij}+{\cal F}_{ij}) = -2i\delta_\kappa E \big[ P_+ \gamma_i E_j
+ P_- \gamma_j E_i] 
\eqn\prelimt
$$
where $P_\pm$ are the following projection operators:
$$
P_\pm = \cases{ {1\over2}(1\pm \Gamma_{11}) & IIA\cr
{1\over2}(1\pm \sigma_3) & IIB }
\eqn\prelimu
$$
We have here supressed the $I,J$ indices on IIB spinors. It follows that
$$
\delta_\kappa {\cal L}_{DBI} =  -i{\cal L}_{DBI}\, 
\delta_\kappa E \big[ (g+{\cal F})^{ij}\gamma_j P_+ 
 + (g-{\cal F})^{ij}\gamma_jP_- \big]E_i 
\eqn\prelimv
$$
where $(g+{\cal F})^{ij}$ is the inverse of $(g+{\cal F})_{ij}$ and we have used
the fact that
$$
(g+ {\cal F})^{ji} = (g-{\cal F})^{ij}\ .
\eqn\prelimw
$$
Note that the $\kappa$-variation of $\phi$ is proportional to the dilatino
(since $\delta_\kappa \phi = \delta_\kappa E^AD_A\phi$ and $\delta_\kappa E^a$
vanishes) and the dilatino vanishes in a bosonic background, by definition.

It will prove convenient to rewrite \prelimv\ in terms of the
$(p+1)\times (p+1)$ matrix
$$
X^i{}_j = g^{ik}{\cal F}_{kj}\ .
\eqn\prelimwa
$$
Thus,
$$
\delta_\kappa {\cal L}_{DBI} = -i{\cal L}_{DBI}\, 
\delta_\kappa E 
\big\{ \big[(1+ X)^{-1}\big]^i{}_j\gamma^j P_+ 
+ \big[(1-X)^{-1}\big]^i{}_j \gamma^j P_-\big\}E_i \ .
\eqn\prelimwb
$$
Using the fact that
$$
(1 \pm X)^{-1} = (1-X^2)^{-1}(1\mp X)
\eqn\prea
$$
we can rewrite \prelimwb\ in the alternative form
$$
\delta_\kappa {\cal L}_{DBI} = -i{\cal L}_{DBI}\,  
\delta_\kappa E N^iE_i
\eqn\dbivar
$$
where
$$
N^i= \cases{\big[(1- X^2)^{-1}\big]^i{}_j \gamma^j 
 + \big[(1- X^2)^{-1}X\big]^i{}_j \gamma^j\, \Gamma_{11} & (IIA) \cr
\big[(1- X^2)^{-1}\big]^i{}_j \gamma^j\otimes {\bf
1}_2 - \big[(1- X^2)^{-1}X\big]^i{}_j \gamma^j\otimes \sigma_3 
& (IIB)}
\eqn\prec
$$

We now turn to the $\kappa$-variation of the WZ term \introg. 
Using \prelimk\ and \prelimn, and omitting a surface term, we find (for a 
bosonic background) that
$$
\delta_\kappa I_{WZ} =\int_W i_{\delta Z}R(C) e^{\cal F} 
\eqn\pred
$$
where $i_{\delta Z}R(C)$ indicates a contraction of the differential form
$R(C)$ with the vector field $\delta_\kappa Z^M$, i.e.
$$
i_{\delta Z}R^{(n)} = {1\over (n-1)!} E^{A_{p+2}} \cdots E^{A_2} \delta
E^{A_1} R^{(n)}_{A_1A_2\dots A_{n}}\ .
\eqn\pree
$$
Using the superspace constraints \twoacon\ and \twobcon, together with the
identity \idengam, we find that
$$
\delta {\cal L}_{WZ} =
\cases{i{\cal L}_{DBI}\, \delta_\kappa E M^i_p
\Gamma_{(0)}E_i & (IIA)\cr
i{\cal L}_{DBI}\, \delta_\kappa E M^i_p\,
i\sigma_2\otimes \Gamma_{(0)}E_i & (IIB) }
\eqn\wzone
$$
where 
$$
M^i_{(p)}= {1\over \sqrt{\det(1+X)}}\sum_{n=0}^\infty {1\over 2^n n!} 
\gamma^{ij_1k_1\dots j_nk_n} X_{j_1k_1}\dots X_{j_nk_n}Q^{(n)}_{(p)}
\eqn\wzthree
$$
with
$$
Q^{(n)}_{(p)} =\cases{(-\Gamma_{11})^{n+ (p-2)/2} & (IIA)\cr
(-\sigma_3)^{n+ (p-3)/2} & (IIB)}
\eqn\wzfour
$$
The infinite sum in \wzthree\ is of course truncated automatically as soon as
$n$ exceeds $p/2$.

Combining the variations of both the DBI and WZ terms in the action we deduce
that
$$
\delta_\kappa I =-\int\! d^{p+1}\xi\, {\cal L}_{DBI}\,
i\delta_\kappa E K^i_{(p)} E_i
\eqn\finala
$$
where 
$$
K^i_{(p)} = N^i + M^i_{(p)}
$$
with $N^i$ and $M^i_{(p)}$ as given in \prec\ and \wzthree, respectively.

We have still to specify the spinor variation $\delta_\kappa E$. On general
grounds it must take the form
$$
\delta_\kappa E = \bar\kappa (1+ \Gamma)
\eqn\finalb
$$
where $\Gamma$ is a matrix with the properties
$$
\Gamma^2=1 \qquad \tr\, \Gamma =0\ .
\eqn\finalc
$$
For $\kappa$-symmetry of the D-p-brane action we require also that
$$
(1+\Gamma)K^i_{(p)}\equiv 0\ .
\eqn\basicid
$$
Given this identity one could deduce that 
$$
K^i_{(p)} =(1-\Gamma)T^i_{(p)}
\eqn\finale
$$
for some matrix $T^i_{(p)}$, which would make the $\kappa$-symmetry
manifest. This was the basis of the strategy for proving $\kappa$-symmetry
proposed in [\schwarz], but since it involves the simultaneous
determination of both $\Gamma$ and $T^i_{(p)}$, we chose instead to
determine $\Gamma$ directly from \basicid. 

Consider the matrix
$$
\Gamma = {1\over \sqrt{\det(1+X)}}\sum_{n=0}^\infty {1\over 2^n n!} 
\gamma^{j_1k_1\dots j_nk_n} X_{j_1k_1}\dots X_{j_nk_n}J^{(n)}_{(p)}
\eqn\finalg
$$
where
$$
J^{(n)}_{(p)} = \cases{(\Gamma_{11})^{n+(p-2)/2} \Gamma_{(0)} & (IIA)\cr
(-1)^n (\sigma_3)^{n+(p-3)/2} i\sigma_2 \otimes \Gamma_{(0)} & (IIB)}
\eqn\finalh
$$
Clearly, $\Gamma$ thus defined has vanishing trace and, for a given value of $p$, standard gamma matrix algebra suffices to establish that $\Gamma^2=1$; we have verified this for $p\le6$. We now claim that this matrix $\Gamma$ is also 
such that \basicid\ is true. To establish this, we begin
by separating those terms on the left hand side of
\basicid\ with a factor of $\sqrt{\det (1+X)}$ from those terms without this
factor. The former cancel straightforwardly, the only subtle point being
that for even $p$ the following identity is needed
$$
X^i{}_{[j}X_{k_1k_2}\cdots X_{k_{p-1}k_{p]}} \equiv 0\ .
\eqn\xiden
$$
The cancellation of the terms {\it without} a factor of $\sqrt{\det (1+X)}$ is
more involved. Firstly, we may separate these terms into four distinct matrix
structures according to whether there is a factor of $\Gamma_{(0)}$ and/or a
factor of $\Gamma_{11}$, in the IIA case, and $\sigma_3$ in the IIB case
(possibly after multiplication by $\sigma_1$ or $\sigma_2$). Each of these
subcalculations involves the reduction of a sum of products of an odd
number of $\gamma$ matrices to the form
$$
\sum_{n=0}^\infty A^i_{(n)}{}_{j_i\dots j_{2k+1}}(X)\, 
\gamma^{j_1\dots j_{2k+1}}
\eqn\finali
$$
where the coefficients $A_{(n)}$ are polynomials in the entries of $X$. The validity of the
identity \basicid\ requires that $A_{(n)}$ vanish unless $n=0$. Using
straightforward gamma-matrix algebra we have verified in detail for $p\le6$ 
that this condition is indeed satisfied. The verification
of $\kappa$-symmetry is thereby reduced to establishing a relation of the
form
$$
\det(1+X)\gamma^i = A^i_{(0)}{}_j(X)\gamma^j
\eqn\finalj
$$
where $A_{(0)}$ is the matrix with entries $A^i_{(0)}{}_j$. Equivalently,
$$
A_{(0)}(X) = \det (1+X) {\bf 1}
\eqn\finalk
$$
where ${\bf I}$ is the $(p+1)\times (p+1)$
identity matrix. One also requires a relation similar to \finalj\ 
in which the left hand side involves
$X^i{}_j\gamma^j$ instead of $\gamma^i$, but this turns out to be an
immediate consequence of \finalj.

The left hand side
of \finalk\ is a polynomial in the matrix $X$, so the validity of the relation
clearly requires that $X$ satisfy some polynomial identity. The identities
satisfied by the $(p+1)\times (p+1)$ matrix $X$ can be obtained as follows. Let
us suppose that $X$ satisfies the identity $P_{p+1}(X)\equiv 0$, where
$P_{p+1}$ is a $(p+1)$th order polynomial; then the $(p+2)\times (p+2)$ matrix
$X$ satisfies the identity $P_{p+2}(X)\equiv 0$, where
$$
P_{p+2}=\cases{P_{p+1}(X)X \qquad p=1,3,5,7,9\cr
P_{p+1}(X)X - {1\over (p+2)}\tr \big( P_{p+1}(X)X\big) \qquad p=2,4,6,8.}
\eqn\finall
$$
Thus all polynomial identities follow from the $p=1$ identity for which
$$
P_2(X) = X^2 -{1\over2}\tr X^2\ .
\eqn\finalm
$$
For example, for $p=4$ one has $P_5(X)\equiv 0$, where
$$
P_5(X) = X^5 -{1\over2} X^3 \tr X^2 +{1\over8} X (\tr X^2)^2 -
{1\over4} X(\tr X^4) \ .
\eqn\finaln
$$
Using these identities, we have verified for $p\le6$ that
$A_{(0)}(X)$ is indeed proportional to the identity matrix, and then that the
coefficient of proportionality is indeed just $\det(1+X)$. For this final step
one needs the expansion of the determinant. Since $p\le9$ this expansion is never needed to beyond the $X^{10}$ level. We record here the expansion to
this level:
$$
\eqalign{
\det(1+X) &= 1 -{1\over2}\tr X^2 - {1\over4}\tr X^4 + {1\over8}(\tr X^2)^2
-{1\over6}\tr X^6 + {1\over8}\tr X^2 \tr X^4 \cr
&-{1\over48}(\tr X^2)^3
-{1\over8}\tr X^8 + {1\over12}\tr X^2\tr X^6 +{1\over32}(\tr X^4)^2 -{1\over32}
\tr X^4 (\tr X^2)^2 \cr
&+ {1\over 4! 2^4}(\tr X^2)^4
-{1\over10}\tr X^{10} + {1\over 16} \tr X^2\tr X^8 + {1\over24}\tr X^4 \tr
X^6\cr 
&-{1\over 48} \tr X^6 (\tr X^2)^2 -{1\over 64} \tr X^2 (\tr X^4)^2 
+ {1\over 4! 2^5}\tr X^4 (\tr X^2)^3 \cr
&-{1\over 5! 2^5} (\tr X^2)^5 + {\cal O}(X^{12})\ .}
\eqn\expdet
$$

%%%%%%%%%%%%%%%%%%%%%%%%%%%%%Chapter 5%%%%%%%%%%%%%%%%%%%%%%%%%%%%

\chapter{The super D-0-brane from D=11}

We now rederive the super D-0-brane action and its $\kappa$-transformations
from D=11. We start with the {\it massless} D=11 superparticle, for
which the action is
$$
I= -\int dt\,  {1\over2\hat v} \hat E_t\cdot \hat E_t 
\eqn\zerod
$$
where $\hat v$ is an independent worldline density and the hats indicate D=11
quantities. This action is invariant under the $\kappa$-transformations
$$
\delta_\kappa \hat E^\alpha = \hat \gamma_t \hat\kappa \qquad \delta_\kappa
{\hat v} = 2i\hat v (\hat\kappa E_t)
\eqn\zeroe
$$
where $\hat\kappa (t)$ is a D=11 spinor parameter and $\hat\gamma_t = \hat
E_t^{\hat a}\Gamma_{\hat a}$. The dimensional reduction to D=10 string-frame fields of the D=11
background fields is achieved by adopting the notation $dx^{\hat m}= (dx^m,dy)$,
$\hat E^{\hat a} = (\hat E^a,\hat E^{11})$ and then taking the supervielbein to
be such that
$$
\eqalign{
\hat E_y{}^a =0 \qquad & \qquad \hat E_y{}^{11} = e^{{2\over3}\phi}\cr
\hat E_M{}^{11} = e^{{2\over3}\phi}C_M \qquad & \qquad
\hat E_M{}^a = e^{-{1\over3}\phi} E_M{}^a\ . }
\eqn\zerof
$$
We also choose
$$
\hat E_M{}^\alpha = E_M{}^\alpha\ .
$$
It then follows that
$$
\eqalign{
\hat E_t{}^a &= e^{-{1\over3}\phi}\, E_t{}^a \cr
\hat E_t{} ^{11} &= e^{{2\over3}\phi}(\dot y + C_t) }
\eqn\zerog
$$
where $C_t= \dot Z^M C_M$. It is convenient to define a new worldline
density $v$ by
$$
v = e^{{2\over3}\phi}\hat v\ .
\eqn\zeroh
$$
With this notation, and in this
Kaluza-Klein (KK) background, the D=11 massless 0-brane Lagrangian is
$$
L= -{1\over2v} E_t^2 -{1\over2v}e^{2\phi} (\dot y + C_t)^2\ .
\eqn\zeroi
$$
We remark that $\delta_\kappa \hat E^{11}=0$ implies that $\delta_\kappa y=
-\delta_\kappa Z^M C_M$, so that $(\dot y +C_t)$ transforms `covariantly'
under $\kappa$-symmetry, i.e. without time derivatives of $\kappa$. The
Euler-Lagrange equations of this Lagrangian allow the solution
$$
(\dot y +C_t) = \mu e^{-2\phi} v
\eqn\zeroj
$$
for arbitrary constant $\mu$ (which has dimensions of mass). This equation can
then be used to eliminate $\dot y$ from the remaining field equations, which
thereby acquire a $\mu$-dependence. It is important to appreciate that these
equations are {\it not} the same as those found by first substituting for
$\dot y$ in \zeroi\ and then varying with respect to $Z^M$. This subtlety has
been addressed elsewhere in a different context [\ant]. It will suffice
here to state that the substitution is permissible if one first adds to the
Lagrangian the total derivative $\mu\dot y$\foot{An alternative procedure is to
rewrite the action in Hamiltonian form and then set $P_y=\mu$, where
$P_y$ is the momentum canonically conjugate to $y$}. One then finds the
new Lagrangian
$$
L =  -{1\over2v} E_t^2 +{1\over2} \mu^2 v e^{-2\phi} -\mu \dot Z^MC_M\ .
\eqn\zerok
$$
Elimination of the auxiliary variable $v$ then yields the action
$$
I = -\mu\int dt \big\{ e^{-\phi}\sqrt{-E_t^2}+ \dot Z^MC_M\big\}\ .
\eqn\zerol
$$
which is the D-0-brane action with mass $\mu$ (set to unity previously) in a
IIA supergravity background with mass parameter $m$ set to zero. The
non-vanishing $\kappa$-variations are
$$
\delta E^\alpha = \delta \hat E^\alpha = \bar\kappa (e^{-{1\over3}\phi}
\gamma_t  + e^{-{4\over3} \phi} \mu v \Gamma_{11})^\alpha \ .
\eqn\zerola
$$
Now $\gamma_t = \sqrt{-E_t^2}\,\Gamma_{(0)}$, and the equation for $v$ is
$\mu v= e^\phi\sqrt{-E_t^2}$, so defining a new parameter $\kappa$ by
$$
\kappa = {1\over \sqrt{-E_t^2}}\, e^{-{1\over3}\phi}\Gamma_{(0)}\hat\kappa\ ,
\eqn\zeron
$$
we find that
$$
\delta E^\alpha = \bar\kappa (1+ \Gamma_{(0)}\Gamma_{11})^\alpha\ ,
\eqn\zerom
$$
which is precisely the $p=0$ case of the $\kappa$-symmetry transformation
derived earlier for the general D-p-brane. 

We learn from this exercise that the D-0-brane action is invariant under the
$\kappa$-symmetry transformation \zerom\ for general backgrounds, i.e. including
fermions, because the D=11 particle action was. This can be checked directly;
imposing the constraint
$$
R^{(2)}_{a\beta} = e^{-\phi} (\Gamma_a\xi)_\beta 
\eqn\fermone
$$
for some spinor $\xi$, one finds that the action \zerol\ is $\kappa$-symmetric 
in a general background if
$$
\chi = -D\phi + \Gamma_{11}\xi\ ,
\eqn\fermtwo
$$ 
where $\chi$ is the fermion field appearing in the torsion constraints \prelimd.
One might have expected $\kappa$-symmetry to fix both $\chi$ and $\xi$ in terms
of the dilatino $D\phi$; the additional freedom arises because of the freedom
in the choice of the `conventional' superspace constraints. The usual choice in
D=11 is $T_{a\beta}{}^c=0$, and this corresponds in D=10 to 
$$
\chi = -{1\over3}D\phi \qquad \xi= {2\over3} \Gamma_{11}D\phi\ .
\eqn\fermthree
$$
Thus, the D-0-brane action is invariant under the $\kappa$-symmetry
transformations given above in a general background of massless IIA
supergravity, since it is the massless IIA theory that is obtained from D=11, 
but the result can be immediately generalized to allow for massive IIA
backgrounds. First, one must include the `CS' term in the D-0-brane action (so
$C^{(1)}\rightarrow C^{(1)} + mV$, where $m$ is the IIA mass parameter). One
then need only impose the same superspace constraints as before but allow for
the $m$-dependent modifications of the R-R field strengths; specifically,
$R^{(2)}$ should now include the term $mB$.

%%%%%%%%%%%%%%%%%%%%%%%%%%Chapter 6%%%%%%%%%%%%%%%%%%%%%%%%%%%%%%%

\chapter{The super D-2-brane from D=11}

We turn now to the derivation of the super D-2-brane action and its
$\kappa$-transformations from D=11. We again use hats to
distinguish D=11 quantities from their D=10 counterparts. The D=11 supermembrane
action has the Lagrangian [\bst]
$$
L = {1\over2\hat v}\det \hat g_{ij} - {1\over2}\hat v + 
{1\over 6} \varepsilon^{ijk} \hat A_{ijk}\ .
\eqn\twoa
$$
We shall use the same KK ansatz as in the previous section, which yields
$$
\eqalign{
\hat E_i^a &= e^{-{1\over3}\phi}E_i^a\cr
\hat E_i^{11} &= e^{{2\over3}\phi} (\partial_i y + C_i) }
\eqn\twob
$$
and hence
$$
\hat g_{ij} = e^{-{2\over3}\phi}\big[g_{ij} + e^{2\phi}(\partial_i y + C_i)
(\partial_j y + C_j)\big]\ .
\eqn\twoc
$$
In addition,
$$
{1\over 6} \varepsilon^{ijk} \hat A_{ijk}= {1\over6} \varepsilon^{ijk} A_{ijk}
+ {1\over2}B_{ij}\partial_k y \ .
\eqn\twod
$$

It will prove convenient to introduce the 1-form field strength
$$
Y= dy +C
\eqn\twoe
$$
which has the Bianchi identity
$$
dY\equiv K\ ,
\eqn\twof
$$
where $K \equiv R^{(2)} = dC$.
We shall also need the identity for $3\times3$ matrices
$$
\det (g_{ij} + e^{2\phi}Y_iY_j) \equiv (\det g) [1+ e^{2\phi}|Y|^2]
\eqn\twog
$$
where $|Y|^2$ is shorthand for $g^{ij}Y_iY_j$.

We can now rewrite the D=11 supermembrane Lagrangian in the KK background as
$$
\eqalign{
L &= {1\over2\hat v} e^{-2\phi}(\det g)[1+ e^{2\phi}|Y|^2] -{1\over2}\hat v\cr
& +{1\over6} \varepsilon^{ijk} (A_{ijk} -3B_{ij}C_k)
+ {1\over2}\varepsilon^{ijk}B_{ij}Y_k \ .}
\eqn\twoh
$$
The strategy will now be to replace the scalar $y$ by its dual, a 1-form gauge potential. To this end we promote the field strength $Y$ to 
the status of an
independent variable, which we can do provided that we add to the action a
Lagrange multipler term for the Bianchi identity \twof, i.e. a term proportional
to $V\wedge (dY-K)$, where $V$ is a 1-form Lagrange multipler field.
Integrating by parts, we see that this is equivalent to
$$
\eqalign{
L_{LM} &= -{1\over2} \varepsilon^{ijk} F_{ij}(Y-C)_k\cr
&= -{1\over2} \varepsilon^{ijk}[ {\cal F}_{ij}(Y-C)_k + B_{ij}(Y-C)_k] }
\eqn\twoi
$$
where we have defined $F=dV$ and ${\cal F}=dV-B$. Adding $L_{LM}$ to the
Lagrangian of \twoh\ we arrive at the new, dual, Lagrangian 
$$
\eqalign{
\tilde L = &{1\over2\hat v} e^{-2\phi}(\det g) -{1\over2}\hat v +{1\over6}
\varepsilon^{ijk} (A_{ijk} +3{\cal F}_{ij}C_k) \cr 
&+ {1\over2\hat v}(\det g) |Y|^2 -{1\over2} \varepsilon^{ijk}{\cal F}_{ij}Y_k\ .}
\eqn\twoh
$$
The one-form $Y$ is now an auxiliary field which may be eliminated by its field
equation
$$
Y^i = {{\hat v}\over 2\det g}\varepsilon^{ijk}{\cal F}_{jk}\ .
\eqn\twoi
$$
Back substitution then yields
$$
\tilde L = {1\over2\hat v} e^{-2\phi}(\det g) -{1\over 2}\hat v (\det g)[1+
|{\cal F}|^2] + L_{WZ}
\eqn\twoj
$$
where
$$
L_{WZ} = {1\over6}\varepsilon^{ijk} (A_{ijk} +3{\cal F}_{ij}C_k)
\eqn\twok
$$
and 
$$
|{\cal F}|^2 \equiv g^{ij} g^{kl}{\cal F}_{ik}{\cal F}_{jl}\ .
\eqn\twol
$$

We now use the further identity for $3\times3$ matrices
$$
(\det g)[1+{1\over2}|{\cal F}|^2) \equiv \det (g+ {\cal F})\ ,
\eqn\twom
$$
and define the new worldvolume density
$$
v= - (\hat v)^{-1}e^{-2\phi} \det g \ ,
\eqn\twon
$$
to rewrite \twoj\ as
$$
\tilde L = {1\over2 v} e^{-2\phi}\det (g +{\cal F}) -{1\over 2}v + L_{WZ}\ .
\eqn\twoo
$$
This new D=10 supermembrane Lagrangian is in a form similar to that with
which we started in D=11. If $v$ is now eliminated by its
Euler-Lagrange equation we obtain the equivalent Lagrangian
$$
L = - e^{-\phi}\sqrt{\det (g +{\cal F})} + L_{WZ}\ ,
\eqn\atwoo
$$
for which the corresponding action is
$$
I= I_{DBI} \ +\ \int\! [ A + {\cal F}\wedge C ]\ .
\eqn\abtwo
$$  
This is precisely the super D-2-brane action given previously with $C^{(2)}=A$
and $C^{(1)}=C$.

As shown in [\pkt], the $\kappa$-symmetry variations of $Z^M$ and $V$ that 
leave this action invariant can also be deduced from its D=11 origin. The
$\kappa$-variation of $Z^M$ is encoded in the D=11 supermembrane
$\kappa$-variation
$$
\delta_\kappa \hat E^\alpha = [\bar\kappa(1+\hat \Gamma)]^\alpha
\eqn\twooa
$$
where $\hat\Gamma$ is the matrix
$$
\hat \Gamma = {1\over6\sqrt{-\det \hat g}}\, \varepsilon^{ijk}\hat E_i^{\hat
a}\hat E_j^{\hat b} \hat E_k^{\hat c}\, \Gamma_{\hat a \hat b\hat c}\ .
\eqn\twoob
$$
When expressed in terms of D=10 variables this becomes
$$
\hat\Gamma= \sqrt{\det (1+X)}\, \Gamma_{(0)} -{1\over2}\gamma^{ij}X_{ij}
\Gamma_{11}
\eqn\twooc
$$
where 
$$
\Gamma_{(0)}= {1\over6\sqrt{-\det g}}\varepsilon^{ijk}\gamma_{ijk}\ .
\eqn\twood
$$
Note that $\hat\Gamma$ has the properties required for $\kappa$-invariance, i.e.
$$
\hat\Gamma^2 =1 \qquad \tr\; \hat\Gamma =0\ .
\eqn\hatgam
$$

The $\kappa$-symmetry variation of $V$, and hence of ${\cal F}$, can be deduced
from D=11 as follows.  The variation of $Y$ in \twoh\ is determined in terms of
the variations $\delta_\kappa Z^M$ via its definition \twoe; so as long as $Y$
satisfies its Bianchi identity $dY=K$ the action $L$ is $\kappa$-invariant.
It then follows, when $Y$ is taken to be an independent variable, that
the $\kappa$-variation of $L$ must have $(dY-K)$ as a factor. Since
$dY-K$ is itself $\kappa$-invariant, the $\kappa$-invariance of $\tilde L$ is
ensured by an appropriate variation of the Lagrange multiplier $V$.
Now, the Bianchi identity $dY=K$ was needed for $\kappa$-invariance of $L$ only to justify the neglect of a term arising after
integration by parts in one term of $\delta_\kappa L_{WZ}$. This term is
easily isolated:
$$
\delta_\kappa L= {1\over2}\varepsilon^{ijk} \delta_\kappa E^A E_i^B B_{BA}
(2\partial _jY_k - K_{jk})\ .
\eqn\twop
$$
This variation of $L$ may be cancelled in $\delta_\kappa\tilde L$ by the
variation
$$
\delta_\kappa V = \delta_\kappa E^A E_i^B B_{BA}\ .
\eqn\twoq
$$
Together with the variation of $B$, given in \preliml\ for general variations
$\delta E^A$, this allows us to determine the $\kappa$-variation of
${\cal F}\equiv dV-B$. The result is
$$
\delta_\kappa {\cal F} = - {1\over2}E^A E^B\delta_\kappa E^C H_{CBA}\ .
\eqn\twor
$$
Observe that the variation of $V$ has simply removed the total derivative
term in the variation of $B-dV$. As we saw earlier, this
result generalizes to all $p$.  We have now recovered the results obtained
earlier for the D-2-brane action, but without the restriction to bosonic
backgrounds, so the $\kappa$-symmetry transformations derived earlier are
equally valid in a general background. We expect that this is equally
true for all $p$.  

A curious feature of the above results is that the matrix $\hat\Gamma$
appearing in the $\kappa$-transformations deduced from D=11 is {\it not} the
same as the specialization to $p=2$ of the matrix $\Gamma$ appearing earlier
in our discussion of the $\kappa$-transformations of the general D-p-brane
action. Thus there are {\it two} matrices satisfying the conditions needed to
establish $kappa$-symmetry! In fact, these two matrices are related as follows:
$$
(1+ \Gamma)= {1\over \sqrt{\det(1+X)}} \Gamma_{(0)}(1+ \hat\Gamma)\ .
\eqn\gamhattwo
$$
Thus, a spinor annihilated by $(1+ \Gamma)$ is also annihilated by
$(1+\hat\Gamma)$, and vice-versa.

%%%%%%%%%%%%%%%%%%%%%%%%%%Chapter 6%%%%%%%%%%%%%%%%%%%%%%%%%%%%%%%

\chapter{Conclusions}

We have presented the super D-p-brane action for all $p\le9$ in general
supergravity backgrounds, including massive supergravity backgrounds in the IIA
case. We have also presented the full $\kappa$-symmetry transformations. We
have fully verified $\kappa$-invariance for bosonic backgrounds for $p\le6$,
and general backgrounds for $p=0,2$. The calculations required for verification
of $\kappa$-symmetry of the D-p-brane action are embedded in those required
for verification of $\kappa$-symmetry of the D-(p+1)-brane, so our results for
$p\le6$ also provide a partial verification of $\kappa$-symmetry for the
remaining $p \ge7$ cases. We have no doubt that $\kappa$-symmetry also holds
in these cases.

We have also shown in detail how the $p=0$ and $p=2$ actions follow from the
massless superparticle and supermembrane actions in D=11. It is believed that
the D-4-brane action should similarly be derivable from the M-theory
super-fivebrane action [\pkt], but the latter is not yet known. We hope that
the results of this paper will provide some clues to the solution of this
outstanding problem. 

\vskip 0.5cm
{\it Acknowledgements}: P.K.T. and E.B. thank each other's institution for
hospitality. The work of E.B. was supported by the European Comission TMR
program ERBFMRX-CT96-0045, in which E.B. is associated to the University of
Utrecht. 
\vskip 0.5 cm
{\bf Note added}: upon completion of this work a paper appeared [\cedertwo] with
similar results to those obtained here. These authors have established
$\kappa$-symmetry for general backgrounds involving fermions but did not
consider massive IIA backgrounds, nor the relation of the D-0-brane to D=11. 
\refout
\end